\RequirePackage{ifpdf}
\ifpdf
  \documentclass[11pt,a4paper]{article}
  \usepackage{type1cm}
  \usepackage[pdftex,colorlinks]{hyperref}
  \usepackage[pdftex]{graphicx}
\else
  \documentclass[11pt,a4paper]{article}
  \usepackage{graphicx}
\fi
\usepackage[small, it]{caption}
\usepackage{thophys}
\usepackage{latexsym} 
\usepackage{amsmath}
\usepackage{amssymb}
\usepackage[english]{babel}
\makeatletter
\newif\if@preliminary
\@preliminaryfalse
\def\preliminary{\@preliminaryfalse}

\renewcommand{\eqref}[1]{(\ref{#1})}

\DeclareMathOperator{\tr}{tr}

\def\preprintno#1{\def\@preprintno{#1}}
\def\address#1{\def\@address{#1}}
\def\email#1#2{\thanks{\tt #1@{}#2}}
\def\abstract#1{\def\@abstract{#1}}
\renewcommand\abstractname{ABSTRACT}
\newlength\preprintnoskip
\newlength\abstractwidth
\@titlepagetrue
\renewcommand\maketitle{\begin{titlepage}
  \let\footnotesize\small
  \hfill\parbox{\preprintnoskip}{
  \begin{flushright}\@preprintno\end{flushright}}\hspace*{1cm}
  \vskip 60\p@
  \begin{center}
    {\Large\bf\boldmath \@title \par}\vskip 1cm
    {\sc\@author \par}\vskip 3mm
    {\@address \par}
    \if@preliminary
      \vskip 2cm {\large\sf PRELIMINARY DRAFT \par \@date}
    \fi
  \end{center}\par
  \@thanks
  \vfill
  \begin{center}
    \parbox{\abstractwidth}{\centerline{\abstractname}
    \vskip 3mm
    \@abstract}
  \end{center}
  \end{titlepage}
  \setcounter{footnote}{0}
  \let\thanks\relax\let\maketitle\relax
  \gdef\@thanks{}\gdef\@author{}\gdef\@address{}
  \gdef\@title{}\gdef\@abstract{}\gdef\@preprintno{}
}

\def\@citex[#1]#2{\if@filesw\immediate\write\@auxout{\string\citation{#2}}\fi
  \def\@citea{}\@cite{\@for\@citeb:=#2\do
    {\@citea\def\@citea{,\penalty\@m}\@ifundefined
       {b@\@citeb}{{\bf ?}\@warning
       {Citation `\@citeb' on page \thepage \space undefined}}%
\hbox{\csname b@\@citeb\endcsname}}}{#1}}
\def\citerange{\@ifnextchar [{\@tempswatrue\@citexr}{\@tempswafalse\@citexr[]}}
\def\@citexr[#1]#2{\if@filesw\immediate\write\@auxout{\string\citation{#2}}\fi
  \def\@citea{}\@cite{\@for\@citeb:=#2\do
    {\@citea\def\@citea{--\penalty\@m}\@ifundefined
       {b@\@citeb}{{\bf ?}\@warning
       {Citation `\@citeb' on page \thepage \space undefined}}%
\hbox{\csname b@\@citeb\endcsname}}}{#1}}
\long\def\@makecaption#1#2{
  \vskip\abovecaptionskip
  \sbox\@tempboxa{#1: \emph{#2}}
  \ifdim \wd\@tempboxa >\hsize
    #1: \emph{#2}\par
  \else
    \hbox to\hsize{\hfil\box\@tempboxa\hfil}
  \fi
  \vskip\belowcaptionskip}

\begin{document}

\preliminary        


\title{A Simplified Scheme for GUT-inspired Theories with Multiple Abelian 
Factors}


\author{
 F.~Braam\email{felix.braam}{physik.uni-freiburg.de}$^{\,a}$
 and J.~Reuter\email{juergen.reuter}{desy.de}$^{\,b}$
}

\address{\it
$^a$Albert-Ludwigs-Universit\"at Freiburg, Physikalisches
  Institut, 
  Hermann-Herder-Str. 3, D--79104 Freiburg, GERMANY
  \\
$^b$DESY Theory Group, Notkestr. 85, D--22603 Hamburg, GERMANY
}

\preprintno{FR-PHENO-2011-011,\\ DESY 11-116}

\abstract{
Grand Unified Theories often involve additional
Abelian group factors apart from the standard model hypercharge, that
generally lead to loop-induced mixing gauge kinetic terms. In this
letter, we show that at the one-loop level this effect can be avoided
in many cases by a 
suitable choice of basis in group space and present a general scheme
for the construction of this basis.\\
In supersymmetric theories however, a residual mixing in the soft SUSY
breaking gaugino mass terms may appear.
We generalize the renormalization group equations for
the gaugino mass terms to account for this effect. In a further
calculation we also present the necessary adjustments in the
renormalization group equations of the trilinear soft breaking
couplings and the soft breaking scalar mass squares.   
}

\maketitle
   
\newpage

\section*{}
The renormalization group equations (RGEs) describe the dependence of
the coupling constants on the choice of the renormalization scale
$\mu$, which is commonly translated into an energy dependence, as the
perturbative series usually converges best if one chooses $\mu$ to be of the
order of the characteristic energy scale of a given process. 
In (supersymmetric) grand unified model
building~\cite{Georgi:1974sy,Pati:1974yy,Fritzsch:1974nn,Achiman:1978vg,Wess:1992cp,Terning:2006bq}  
these equations constitute the 
framework which is employed to derive the potential unification of
the gauge interactions into one fundamental force. They also
describe the evolution of all other Lagrangian parameters (including
the soft supersymmetry breaking parameters mediated to the ``visible''
sector through some mechanism at high scales) from a high unification
scale down to the energy scales accessible to current collider
experiments. 

The RGEs for a (supersymmetric) model with an arbitrary semi-simple
gauge group augmented by at most one $U(1)$ gauge group were given in
\cite{Jones:1974mm,Machacek:1984zw,Machacek:1983fi,Machacek:1983tz}
(\cite{Jones:1974pg,Jones:1983vk,Martin:1993zk,Martin:1993zk}).  
We present a  way of treating the case with several Abelian gauge groups,
including a consistent generalization of the one-loop SUSY RGEs from
\cite{Martin:1993zk} in the case, where a mixing of gauge kinetic terms at
the tree-level does not occur, i.e.
\begin{equation}
  \kappa F^{i}_{\mu\nu}\, F^{\mu\nu, j} \;=\;0 \hspace{1.5cm} \forall
  i\neq j\,.  
  \label{crit0}
\end{equation} 
In this situation we will show that the general concept presented in 
\cite{delAguila:1988,Ferroglia:2006mj} which has recently been applied
to the complete set of SUSY two-loop RGEs in \cite{Fonseca:2011vn}, 
where an additional coupling
parametrizing the mixing is introduced, can be simplified considerably
at the one-loop level
by an appropriate choice of basis for the $U(1)$ groups.


First, we specify the type of models in which condition \eqref{crit0}
holds, i.e. where our formalism is applicable: Consider some
potentially multi-scale symmetry breaking scenario:
\begin{equation}
  \mathcal{G}^{(0)}_N\,
  \stackrel{\Lambda}{\rightarrow}\,
  \left.\mathcal{G}^{(1)}_{N-d}\right|_{d\ge3}\times U(1)^{d-1}\,\rightarrow
  \mbox{SM},
  \label{breaking}
\end{equation}
where $\mathcal{G}^{(0)}_N$ denotes a simple%
\footnote{$\mathcal{G}^{(0)}_N$ does not necessarily have to be simple
  as long as the particle content can be assembled into complete
  multiplets of a simple Lie group and there are no mixing gauge
  kinetic terms at the tree-level.}  
Lie group of rank $N$
and $\mathcal{G}^{(i)}_{N^\prime}$ an arbitrary semi-simple non-Abelian
subgroup of it. 
This implies, that condition \eqref{crit0} holds at $\Lambda$ for
all  $U(1)$ groups in $\mathcal{G}^{(1)}_{N-d}\times
U(1)^2$,  as they all originate from  non-Abelian gauge
multiplets  above $\Lambda$. A term as in Eq.~\eqref{crit0} would have
to necessarily arise from a matching condition at scale $\Lambda$ 
\begin{equation}
  \kappa G^{\phantom{2}}_{\mu\nu,1}\, G^{\mu\nu}_{2} \rightarrow
  \kappa^\prime F^{\phantom{2}}_{\mu\nu,1}\, F^{\mu\nu}_{2} \; , 
\label{imposs}
\end{equation} 
with $G_{\mu\nu}$ being a non-Abelian field-strengh
tensor, which istself is not gauge invariant. So the left-hand
side of Eq.~\eqref{imposs} is forbidden by the gauge symmetry.
 
Our argumentation holds if there are intermediate symmetry breaking
steps above $\Lambda$, with arbitrary semi-simple gauge groups, as
long as the rank $N$ is preserved and the matter content still fills
complete multiplets of $\mathcal{G}^{(0)}_N$. 
At the tree-level, there cannot
appear mixing terms in the course of symmetry breaking, for the same
reason as in Eq.~\eqref{imposs}.  
Furthermore the one-loop corrections to $\kappa^\prime$ in 
any intermediate phase vanish, as 
the
trace over complete representations of the full GUT group vanishes for
products of different generators:%
\begin{equation*}
  \tr[T^AT^B] \;=\;0 \hspace{1.5cm} \forall \;A\neq B. 
\end{equation*}
At the breaking scale $\Lambda$,
a mixing among the $U(1)$ gauge-kinetic terms may be
induced via quantum corrections from matter of representations made
incomplete by the symmetry breaking. Typical examples for such
scenarios can be found in~\cite{Braam:2010sy,Kilian:2006hh} and arise
e.g. in GUT breaking chains like $E_6 \to SO(10)\times U(1)$ or $E_6
\to SU(5) \times U(1)^2$. 
  
We will now develop a scheme for constructing the basis for the $U(1)$
gauge groups, such that the mixing induced at the one-loop level can
be avoided in the $\left.\mathcal{G}^{(3)}_{N-d}\right|_{d\ge3}\times
U(1)^{d-1}$ phase. 

At $\Lambda$, we demand the gauge covariant
derivative to be continuous:
\begin{equation}
  \left.D^i_\mu\,\right|_{\Lambda} \; 
  = \; 
  \left.D^{i+1}_\mu\,\right|_{\Lambda}.
\end{equation}  
From this we obtain a system of linear equations for the couplings and
the charges of the new gauge group as functions of the corresponding
parameters of the mother group:
\begin{equation}
\begin{array}{rllll}
g^\prime_1\, Q^\prime_1 &  =\;  \lambda_{1,1}\; g_1\, Q_1  & +\;
\ldots   & +\;\; 
\lambda_{1,d}\; g_d\, Q_d ,\\
\vdots\ \  &  &  & \vdots &  \\
g^\prime_{d-1}\,Q^\prime_{d-1}& =  \; \lambda_{d-1,1}\; g_1\,
Q_1& + \; \ldots   & +\;\; \lambda_{d-1,d}\; g_d\,
Q_d , \\
g^{\prime}_d\,Q^{\prime}_d &=\;  \lambda_{d,1}\; g_1\, Q_1  & +\;\ldots &+\;\;
\lambda_{d,d} \;g_d\, Q_d.
\end{array}
\label{lin_combo}
\end{equation} 
Here, $U(1)^\prime_i \;i=1,\ldots,d-1 $ are the remaining
unbroken groups in 
$\left.\mathcal{G}^{(3)}_{N-d}\right|_{d\ge3}\times U(1)^{d-1}$, and
$U(1)^{\prime}_d$ corresponds to the broken Cartan generator of the
non-Abelian gauge group, respectively. In Eq.~\eqref{lin_combo} there
are $d^2 +2d$ free parameters $\{g^\prime_i, Q^\prime_i,
\lambda_{i,j}\}$ which can be uniquely determined (up to signs) by
applying the following 
constraints:  
\begin{enumerate}
 \item The vacuum expectation value breaking the symmetry at scale
 $\Lambda$ is not charged under the unbroken $U(1)$ groups:
  \begin{equation*}
   Q^\prime_i \, \braket{ H_\Lambda}
   \;=\;0 
   \hspace{1.5cm} \text{for} \; i=1,\ldots,d-1;
  \end{equation*}
 \item The broken and unbroken generators are normalized according to
   the Dynkin index of some complete GUT representation $R$, $S(R)$ (no
   summation over $i$):
  \begin{equation*}
    \sum_{\alpha\in R} \;Q^\prime_i(\alpha)Q^\prime_i(\alpha)
    \; =\; S(R) \,
    \hspace{.7cm} 
    \text{for} \; i=1,\ldots,d;
  \end{equation*}
 \item Vanishing mixing at the one-loop level:
  \begin{align*}
    0\;=&\;\frac{1}{3}\sum_{R^\phi} \sum_{\alpha\in R^\phi} Q^\prime_i(\alpha)
    Q^\prime_j(\alpha) \,
    +\,\frac{2}{3}\sum_{R^\psi} \sum_{\beta\in R^\psi} Q^\prime_i(\beta)
    Q^\prime_j(\beta) 
  \\
    &\text{for} \; i\neq j; \;i=1,\ldots,d-1;\ \ j=1,\ldots,d-1,
  \end{align*}
  where $R^\phi$ and $R^\psi$ denote representations of scalars
  and Weyl fermions, respectively.  
 \item Corresponding transformations of the gauge fields are
   orthogonal:  
  \begin{equation}
     (A^{\prime\mu}_{1},\ldots,A^{\prime\mu}_d)\;=\; 
    (A^{\mu}_{1},\ldots,A^{\mu}_d)\mathbf{\lambda}^T, 
    \hspace{1.5cm}\mbox{with}
    \hspace{.7cm }
    \lambda_{ij}\,\lambda_{kj} \,=\,\delta_{ik};
    \label{constraints}
  \end{equation}
\end{enumerate} 
A suitable framework to construct the charge operators in
\eqref{lin_combo} and \eqref{constraints} is the concept of 
projection matrices in the weight space of the gauge groups, as
presented in \cite{Slansky:1981yr}. The representations to sum over in
the constraints 2. and 3. as well as the field with non-trivial vacuum
expectation value breaking the
symmetry in constraint 1. can then be represented by their
corresponding weights.    

Applying this procedure at the scale where the rank of the
gauge group is reduced for the first time in the chain of symmetry
breakings ensures that there does not occur any mixing among
gauge-kinetic terms at the one-loop level. If there are subsequent
symmetry breaking mechanisms leading again to several Abelian gauge
groups, the procedure can be applied repeatedly.
The advantage of this scheme is, that one can still apply
the one-loop RGEs as given in \cite{Jones:1974mm,Machacek:1984zw,Machacek:1983fi,Machacek:1983tz}
~\cite{Martin:1993zk} for gauge and superpotential 
couplings without any changes. 

Although, by this procedure one can avoid a mixing in the gauge 
kinetic term, a remnant shows up for softly broken supersymmetric
models, unless the corresponding gaugino masses at the breaking scale are
degenerate, as
\begin{equation*}
  \mathcal{L}_{M_g}\,=\, \bar{\tilde{\chi}}_i M_{ii} \tilde{\chi}_i 
                 \,=\, \bar{\tilde{\chi}}^a 
                 \underbrace{\lambda^{a}_{i}M_{ij}\lambda^{b}_{j}}
		 _{\equiv M^{ab}} \tilde{\chi}^b \hspace{.7cm}                 
\Rightarrow \;\hspace{.7cm} M^{ab} \neq \mbox{diag, unless} \; M_{ii} \equiv M
                 \;\forall i. 
\label{gaugino_mass}
\end{equation*}  

At first glance, non-degenerate gaugino masses seem rather artificial
(though the SUSY-breaking mechanism might not be completely
$U(1)$-blind). Note however, that they naturally appear in the
multi-scale models mentioned above, as $U(1)$ gaugino masses can
evolve differently between intermediate scale above $\Lambda$, as well
as in SUSY-breaking mechanisms sensitive to the beta function of the
gauge group under consideration like AMSB~\cite{amsb}. Another
example where non-degenerate gaugino mass terms
could arise are mixed mediation mechanisms like e.g. mirage
mediation~\cite{mirage}.  

In the following, we will present  the generalized RGEs in the
$\overline{\mbox{DR}}$ scheme~\cite{drbar} at the one-loop 
level for soft SUSY breaking terms, accounting for this effect. We use 
the conventions and nomenclature of~\cite{Martin:1993zk}.     
In the absence of $U(1)$ mixing the one-loop RGEs for the
soft-breaking gaugino masses are given by~\cite{Martin:1993zk}
\begin{equation}
  \frac{d}{dt}M_a\,=\,\frac{2}{16\pi^2}\,\beta_a \, g^2_a\,M_a, 
  \label{rge_gino0}
\end{equation}
with
\begin{equation}
  \beta_a \,=\, S(R)\, -\, 3C(G), 
  \label{betafunk}
\end{equation}  
where $S(R)$ is the Dynkin index summed over all chiral superfields,
and $C(G)$ the quadratic casimir of the adjoint representation,
respectively. The logarithm of the ratio of scales is denoted as $t =
\log \mu/\mu_0$. In the presence of gaugino mass mixing terms, we have
to extend Eq.~\eqref{rge_gino0} in order to describe the running of
the full gaugino mass matrix:
\begin{equation}
  \frac{d}{dt}M_{ab}\,=\,\frac{1}{16\pi^2}\left(\beta_a \, g^2_a\,M_{ab}
  \,+\,                       \beta_b \, g^2_b\,M_{ab}\right). 
\label{rge_gino}
\end{equation}
These contributions arise from the following diagrams
\begin{align*}
  \includegraphics{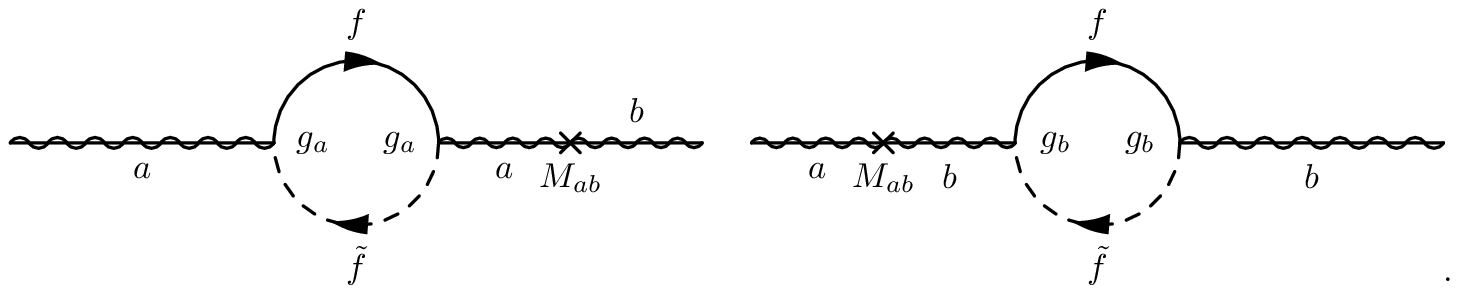}
\end{align*}
Note, that the running of the diagonal entries in the gaugino mass
matrix are not altered in the presence of mixed mass terms, as
diagrams like   
\begin{equation*}
  \includegraphics{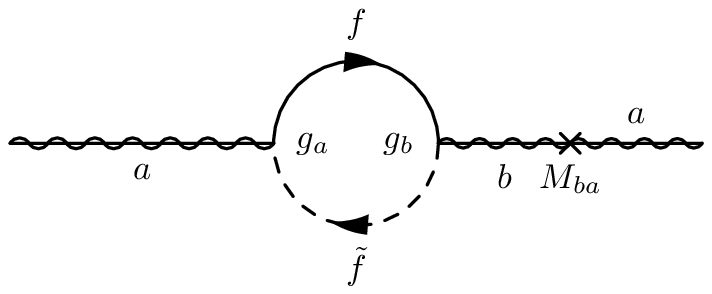}
\label{diag_contrib}
\end{equation*}
are by construction canceled out after rotating to the new $U(1)$
basis.

Besides this effect on the running of the full gaugino mass matrix,
there is also a modification of the RGEs for the trilinear soft
breaking parameters by non-diagonal gaugino mass terms. The new
diagrams contributing to the running of the trilinear terms are of the
form:  
\begin{align*}
  \includegraphics{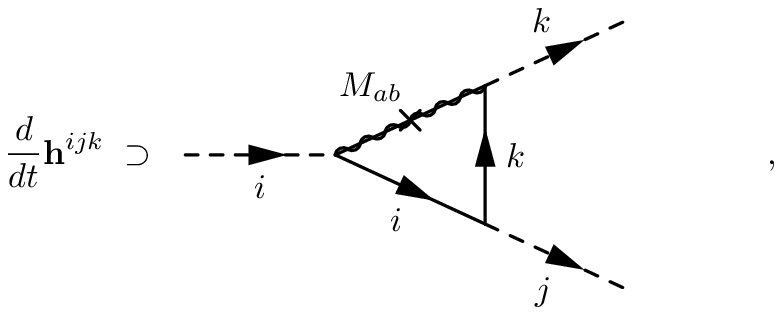}
      \notag 
\end{align*}
The corresponding renormalization group equations for the soft
trilinear terms now read
\begin{align}
  \frac{d\mathbf{h}^{ijk}}{dt}\;=\;\frac{1}{16\pi^2}\Bigl[\,
    \frac{1}{2}&\mathbf{h}^{ijl}\mathbf{Y}_{lpq}\mathbf{Y}^{pqk}
    \,+\,\mathbf{Y}^{ijl}\mathbf{Y}_{lpq}\mathbf{h}^{pqk}
    +
    2\left(\delta_{ab}\mathbf{h}^{ijk}\,-\,2M_{ab}\mathbf{Y}^{ijk}\right)g_ag_b
    C_{ab}(k) 
    \Bigr] \notag \\ &
    \quad + (k \leftrightarrow i) + (k \leftrightarrow j),
\label{rge_h}
\end{align}
with
\begin{equation*}
  C_{ab}(k)=\Bigl[\begin{array}{c}
                            C_a(k) \hspace{1cm} \mbox{if}\hspace{.6cm}
                            a=b\\                                  
			    Q^a_k Q^b_k \hspace{1cm} \mbox{if}
                            \hspace{.6cm} a\neq b                                  
			    \end{array}\Bigr.. 
\end{equation*}
The first line has to be applied for non-Abelian groups, while the
second line accounts for  $U(1)$ factors. 

Finally, for the scalar mass squared terms, the generalization of the 
RGE result calculated in~\cite{Martin:1993zk}, adjusted to take $U(1)$
mixing into account, reads:
\begin{align}
  \frac{d}{dt}(\mathbf{m}^2)^i_j\;=\;\frac{1}{16\pi^2}\Bigl[\;
   & \frac{1}{2} \mathbf{Y}_{jpq}\mathbf{Y}^{pqn}(\mathbf{m}^2)^i_n
    \; 
    \;+\;
     \frac{1}{2}
     \mathbf{Y}^{ipq}\mathbf{Y}_{pqn}(\mathbf{m}^2)^n_j  
    \;+\; 
     2 \mathbf{Y}_{jpq}\mathbf{Y}^{ipn}(\mathbf{m}^2)^q_n
    \; + \mathbf{h}_{jpq}\mathbf{h}^{ipq}
    \notag \\ & \;
    \;-\;
     8\delta^i_j |M_{ab}|^2 g_ag_b\mathcal{C}_{ab}(j) 
     \;+\;
     2g^2_\beta \delta^i_j Q^\beta_j \left(\delta^l_k Q^\beta_k
   (\mathbf{m}^2)^k_l  \right)\Bigr], 
\label{rge_m2}
\end{align}
where in the last term the index $\beta$ runs over all $U(1)$
factors. 

\vspace{1cm}

\textit{Conclusions:} \hspace{0.5cm}

In this letter, we studied a class of (supersymmetric) GUT models,
where $U(1)$ mixing forbidden at tree-level can occur at the
(one-)loop level and defined a generic matching scheme for the
couplings at intermediate thresholds by constructing a suitable choice
of basis. Such a basis avoids -- at the one-loop level -- the mixing
of the gauge couplings and 
completely diagonalizes the $U(1)$ vector (super-) fields (in the absence
of gaugino mass non-degeneracy). In that case, using this specific
basis allows to use the setup in~\cite{Machacek:1984zw,Martin:1993zk}
without any 
changes. In phenomenologically interesting SUSY GUT models, however,
gaugino masses at some high or intermediate scale could be
non-degenerate, mostly by means of running or through some explicit
construction. For such a case, gaugino masses can not longer be
simultaneously diagonalized. Consequently, we gave the modifications
for the renormalization group equations for the gaugino masses, the
trilinear soft breaking terms and the sfermion soft mass squared terms
at the one-loop order.


\subsubsection*{Acknowledgements}

We thank L.~Basso, S.~Dittmaier, C.~Horst, F.~J\"order, A.~Knochel, F.~Staub, 
and J.J.~van der
Bij for stimulating remarks and discussions. This work has been
supported by the German Research Council (DFG) under Grant
No. RE/2850/1-1 as well as by the Ministery for Research and Culture
(MWK) of the German state Baden-W\"urttemberg, and has also been
partially supported by the DFG Graduiertenkolleg GRK 1102 ``Physics at
Hadron Colliders''.

\vspace{5mm}
       
       {\em Note added:}
       As this paper has been finished there has been a similar work
       calculating the 1- and 2-loop RGEs for two mixing $U(1)$s in a
       non-diagonal basis, \cite{Fonseca:2011vn}.



\baselineskip15pt

\begin{thebibliography}{19}
\bibitem{Georgi:1974sy}
  H.~Georgi, S.~L.~Glashow,
  Phys.\ Rev.\ Lett.\  {\bf 32}, 438-441 (1974).

\bibitem{Pati:1974yy}
  J.~C.~Pati, A.~Salam,
  Phys.\ Rev.\  {\bf D10}, 275-289 (1974).

\bibitem{Fritzsch:1974nn}
  H.~Fritzsch, P.~Minkowski,
  Annals Phys.\  {\bf 93}, 193-266 (1975).
  
\bibitem{Achiman:1978vg}
  Y.~Achiman, B.~Stech,
  Phys.\ Lett.\  {\bf B77}, 389 (1978).


\bibitem{Wess:1992cp}
  J.~Wess, J.~Bagger,
  ``Supersymmetry and supergravity,''
  Princeton, USA: Univ. Pr. (1992) 259 p.

\bibitem{Terning:2006bq}
  J.~Terning,
  ``Modern supersymmetry: Dynamics and duality,''
  Oxford, UK: Clarendon (2006) 324 p.

\bibitem{Jones:1974mm}
  D.~R.~T.~Jones,
  Nucl.\ Phys.\  {\bf B75}, 531 (1974).
  
\bibitem{Machacek:1984zw}
  M.~E.~Machacek, M.~T.~Vaughn,
Theory. 3. Scalar Quartic Couplings,''
  Nucl.\ Phys.\  {\bf B249}, 70 (1985).

\bibitem{Machacek:1983fi}
  M.~E.~Machacek, M.~T.~Vaughn,
Theory. 2. Yukawa Couplings,''
  Nucl.\ Phys.\  {\bf B236}, 221 (1984).

\bibitem{Machacek:1983tz}
  M.~E.~Machacek, M.~T.~Vaughn,
Theory. 1. Wave Function Renormalization,''
  Nucl.\ Phys.\  {\bf B222}, 83 (1983).

\bibitem{Jones:1974pg}
  D.~R.~T.~Jones,
Approximation,''
  Nucl.\ Phys.\  {\bf B87}, 127 (1975).

\bibitem{Jones:1983vk}
  D.~R.~T.~Jones, L.~Mezincescu,
  Phys.\ Lett.\  {\bf B136}, 242 (1984).

\bibitem{Martin:1993zk}
  S.~P.~Martin and M.~T.~Vaughn,
  Phys.\ Rev.\  D {\bf 50}, 2282 (1994)
  [Erratum-ibid.\  D {\bf 78}, 039903 (2008)]
  [arXiv:hep-ph/9311340].

\bibitem{delAguila:1988}
  F.~del Aguila, G.~D.~Coughlan and M.~Quiros,
  Nucl.\ Phys.\  B {\bf 307} (1988) 633
  [Erratum-ibid.\  B {\bf 312} (1989) 751].

\bibitem{Ferroglia:2006mj}
  A.~Ferroglia, A.~Lorca and J.~J.~van der Bij,
  Annalen Phys.\  {\bf 16} (2007) 563
  [arXiv:hep-ph/0611174].
\bibitem{Fonseca:2011vn}
  R.~Fonseca, M.~Malinsky, W.~Porod, F.~Staub,
  [arXiv:1107.2670 [hep-ph]].
 
\bibitem{Braam:2010sy}
  F.~Braam, A.~Knochel and J.~Reuter,
  JHEP {\bf 1006}, 013 (2010),
  [arXiv:1001.4074 [hep-ph]].

\bibitem{Kilian:2006hh}
  W.~Kilian, J.~Reuter,
  Phys.\ Lett.\  {\bf B642}, 81-84 (2006).
  [hep-ph/0606277].


\bibitem{Slansky:1981yr}
  R.~Slansky,
  Phys.\ Rept.\  {\bf 79} (1981) 1.

\bibitem{amsb}
  L.~Randall, R.~Sundrum,
  Nucl.\ Phys.\  {\bf B557}, 79-118 (1999).
  [hep-th/9810155];
  G.~F.~Giudice, M.~A.~Luty, H.~Murayama, R.~Rattazzi,
  JHEP {\bf 9812}, 027 (1998).
  [hep-ph/9810442].

\bibitem{mirage} 
  S.~Kachru, R.~Kallosh, A.~D.~Linde, S.~P.~Trivedi,
  Phys.\ Rev.\  {\bf D68}, 046005 (2003).
  [hep-th/0301240];
  K.~Choi, A.~Falkowski, H.~P.~Nilles, M.~Olechowski, S.~Pokorski,
  JHEP {\bf 0411}, 076 (2004).
  [hep-th/0411066].

\bibitem{drbar}
  W.~Siegel,
  Phys.\ Lett.\  {\bf B84}, 193 (1979);
  D.~M.~Capper, D.~R.~T.~Jones, P.~van Nieuwenhuizen,
  Nucl.\ Phys.\  {\bf B167}, 479 (1980);
  J.~A.~Aguilar-Saavedra {\it et al.},
  Eur.\ Phys.\ J.\  {\bf C46}, 43-60 (2006).
  [hep-ph/0511344].
  








\end{thebibliography}
\end{document}